# Sparse Integer Fourier Transform and Its Application in Machine Learning

Jinyang Li[1], Jie Chen[1*]

[1]Department of Mechanical Engineering, Virginia Polytechnic Institute and State University, Blacksburg, VA 24061, USA

## Abstract

The Discrete Fourier transform, mainly implemented by Fast Fourier Transform (FFT), has been essential for numerous domains, such as digital signal processing and feature extraction in materials design. However, current FFT methods suffer from extensive non-trivial multiplications and twiddle-factor operations in the big data era. To improve the computational efficiency of FFT, this study proposes Sparse Integer Fourier Transform (SIFT), a method that leverages the fast computing of sparsity and small integers, and applies the Radix-n method to derive an FFT. The $2^N$-point DFT matrix in Radix-n FFT is decomposed into four sparse integer matrices. A diagonal matrix containing complex twiddle factors is efficiently approximated using an optimization strategy while the remaining matrices retain sparsity and integer entries. For radix-32 FFT, SIFT reduces the estimated arithmetic cost by 75%, with RMSE below $2.3765\times10^{-8}$ and MAPE below 0.01089%. We evaluate frequency extraction in a numerical test and an engineering case. In the numerical test, MAPE and RMSE are below 2.016% and $2.536\times10^{-6}$, respectively. For wind turbine fault diagnosis, SIFT extracts compact spectral features that retain fault information while reducing the classifier's input dimensionality relative to the full time series.

[1] *Corresponding author.

*E-mail address*: jiechen@vt.edu (J. Chen)

# 1. Introduction

Fourier Transform, a method transforming the signal from the time domain to the frequency domain [1], is a widely used method in manifold scenarios, such as signal processing [2, 3], numerical analysis [4, 5], vibration detection and fault diagnosis of mechanical systems [6, 7], pattern recognition [8], and image denoising [9]. However, the Discrete Fourier Transform (DFT) is limited due to its large computational complexity of $O(N^2)$ where $N$ is the number of samples. One of the most widely used algorithms of the Fourier Transform is the Fast Fourier Transform (FFT), which reduces the computational cost of the $N$-point DFT matrix from $O(N^2)$ to $O(Nlog_2N)$ by leveraging the periodicity and symmetry properties of the $N^{th}$ root of unity $e^{-\frac{j2\pi kn}{N}}$ [10]. Built on the foundation of the FFT, some other methods were developed, such as the Split Radix Algorithm (Radix-$n$), Sparse Fourier Transform, and Integer Fourier Transform.

Although current FFT methods have significantly accelerated Fourier Transform, their computational costs are still high and limit their applications to modern data processing problems. For example, digital twins, when online optimization and control are involved, require real-time sensor data processing [11]. Therefore, a FFT method is highly desired that can further improve computing efficiency. Approximate DFT (ADFT) [12] is a promising proposal to the above challenges. The key idea of ADFT is to leverage the approximation of DFT matrices to accelerate the FFT process. A bottleneck that limits the FFT computing efficiency is the operation of complex numbers. Integer elements, such as $0, \pm 1, \pm j$, use shift-add forms, which are more efficient than real multiplications [13-15]. Not only can using simple elements reduce computational cost, but limiting the number of these elements to relatively low integers, such as $\{0, \pm 2^n, \pm j2^n \mid n = 1, 2, 3\}$, is also a promising approach to improve accuracy while maintaining a balance with computational efficiency [16-18]. This approach enables the generation of structured and hardware-friendly matrices suitable for real-time applications, while retaining acceptable accuracy for many engineering problems.

ADFT introduces matrix decomposition, leading to sparse and integer matrices, which reduces the complexity of multiplications, and the number of multiplications is mitigated by the sparsity of matrices. ADFT sacrifices accuracy as the matrix elements cannot be represented precisely by integer elements.

To further accelerate ADFT-based FFT and ensure high approximation accuracy, we proposed a novel ADFT method, Sparse Integer Fourier Transform (SIFT). SIFT reduces computational cost by using sparse and integer matrices for the approximation of the DFT matrix, and achieves minimal approximation error using optimization algorithms to under sparsity and integer constraints. Firstly, the $2^N$-point DFT matrix is rearranged by multiplying by a permutation matrix. After that, the DFT matrix is decomposed into three sparse integer matrices. When approximating the DFT matrices, we first try to use the integer elements in sparse matrices as much as possible. Then, a few elements ($2^{N-1}$-1) of the decomposed matrix are approximated by complex numbers made by integers. Thus, only small round-off errors occur during DFT approximation, while the processing rate increases due to the integer elements and sparsity of the decomposed matrices. The proposed method is used to approximate radix-2, radix-4, radix-8, radix-16, and radix-32 DFT matrices. Theoretical results and analysis show SIFT outperforms other state-of-the-art ADFT methods in terms of accuracy and efficiency. SIFT is tested in two numerical problems and is used for data processing for classification in machine learning in an engineering problem.

As a summary, this paper has the following contributions:

(1) A sparse-and-integer approximation framework is proposed, which replaces most of the expensive floating-point complex multiplications in a Fourier transform with simple operations, such as additions, sign-change operations, and bit shifts.

(2) Instead of storing an entire dense Fourier matrix, the proposed method keeps the transformation in a factorized form with the simplest twiddle factors for reconstruction. Therefore, there is no need to store the full Fourier matrix, and the implementation becomes lighter and easier.

(3) The proposed approximation preserves the FFT behavior with marginal reconstruction error, which accelerates frequency-domain analysis in engineering and data-processing pipelines without sacrificing accuracy.

The paper is organized as follows. Section 2 reviews the related work of the Fourier Transform. The proposed method SIFT used for radix-2, radix-4, radix-8, radix-16, and radix-32 matrices of the FFT process is presented in Section 3. In Section 4, we evaluate the performance proposed method. In Section 5, there are a synthetic function and a numerical problem to test the accuracy of the proposed method. Section 6 applies it to a fault diagnosis case. Section 7 summarizes our conclusions.

# 2. Related Work

## 2.1 Overview Discrete Fourier Transform (DFT)

In general, the $N$-point Fourier Transform is

$$X^F(k)=\sum_{n=0}^{N-1} x(n)W_N^{kn}, k=0,1,\dots,N-1, \tag{1}$$

where $W_N^{kn}=e^{-\frac{j2\pi kn}{N}}$ is the primitive $N^{\text{th}}$ root of unity. $x(n)$ is the $n^{th}$ point of the original signal, $e$ is the natural constant, $N$ is the number of points, and $X^F(k)$ represents the $k^{th}$ point of the DFT result, i.e., the frequency at the $k^{th}$ bin. The $2^N$-point DFT matrix can be represented by

$$\mathbf{F}_{2^N}=\begin{bmatrix} 1 & 1 & \cdots & 1 \\ 1 & W_{2^N}^{1} & \cdots & W_{2^N}^{2^N-1} \\ \vdots & \vdots & \ddots & \vdots \\ 1 & W_{2^N}^{2^N-1} & \cdots & W_{2^N}^{(2^N-1)(2^N-1)} \end{bmatrix}. \tag{2}$$

For a $2^N$-point signal, the FFT response can be calculated by

$$\mathbf{X}^F=\mathbf{F}_{2^N}\mathbf{x} \tag{3}$$

where $\mathbf{X}^F$ is a $2^N\times 1$ vector representing the DFT output, and $\mathbf{x}$ is the original signal ($2^N\times 1$ vector). The longer of the signal, the higher of the computational cost that DFT requires, as the DFT matrix $\mathbf{F}_{2^N}$ needs to calculate $2^N\times 2^N$ twiddle factors and then process the matrix multiplication with the signal to complete the Fourier Transform.

FFT and SFT have been developed to mitigate the computing challenges of DFT, and are introduced in Section 2.2 and 2.3, respectively. Recent research in Fourier Transform algorithms has focused on improving efficiency and accuracy. Especially, new algorithms aim to further exploit the sparsity and structure of signals, leading to even lower computational complexity [19].

### 2.2 Fast Fourier Transform (FFT)

Proposed by Cooley and Tukey half a century ago, FFT enables computing the Fourier transform rapidly by factorizing the DFT matrix into a product of sparse (mostly zero) factors. Therefore, FFT has been widely used in many applications, such as science, engineering, music, and mathematics, and has been described as one of the top 10 Algorithms of the 20$^{th}$ Century [20]. The remainder of this subsection introduces a representative implementation of the FFT algorithm.

#### 2.2.1 Split Radix Algorithm (Radix-n)

One implementation of the FFT algorithm is the Radix-$n$ method (Fig. 1). The Radix-$n$ method utilizes a divide-and-conquer strategy, which determines how FFT splits the DFT calculation task. In every single task, the time series of a signal is processed by the radix-$n$ matrices. For radix-$n$ FFT in Eq. (1), it requires $N=n^m$ points ($m$ should be an integer or after zero-padding to the nearest such length). Therefore, the input signal can be divided into integer groups with $n$ data points and then process the simplest FFT by radix-$n$ matrices in Eq. (2) and Eq. (3), also shown in Fig. 1. Commonly used algorithms are radix-2, radix-4, radix-8, radix-16, and radix-32 FFT, which require the length of the signal to be $N=n^m$ (zero-padding can be used to extend the signal length to meet the FFT computation requirements when the signal length is less than the requirements), respectively. Compared to the radix-2 method, the radix-4 method is more efficient as it reduces the number of multiplications. Furthermore, higher radices (e.g., radix-8/16/32) reduce the number of FFT stages from $\log_2 N$ to $\log_n N$, which lowers the total cost. However, they are more complex to implement because each butterfly is larger and requires more distinct twiddle factors, leading to more complicated computations compared with radix-2 and radix-4. As shown in Fig. 1, the radix-8 matrix is significantly more complex than the radix-2 and radix-4 matrices. Moreover, due to space limitations, the explicit forms of the radix-16 and higher matrices and their corresponding butterfly diagrams are omitted.

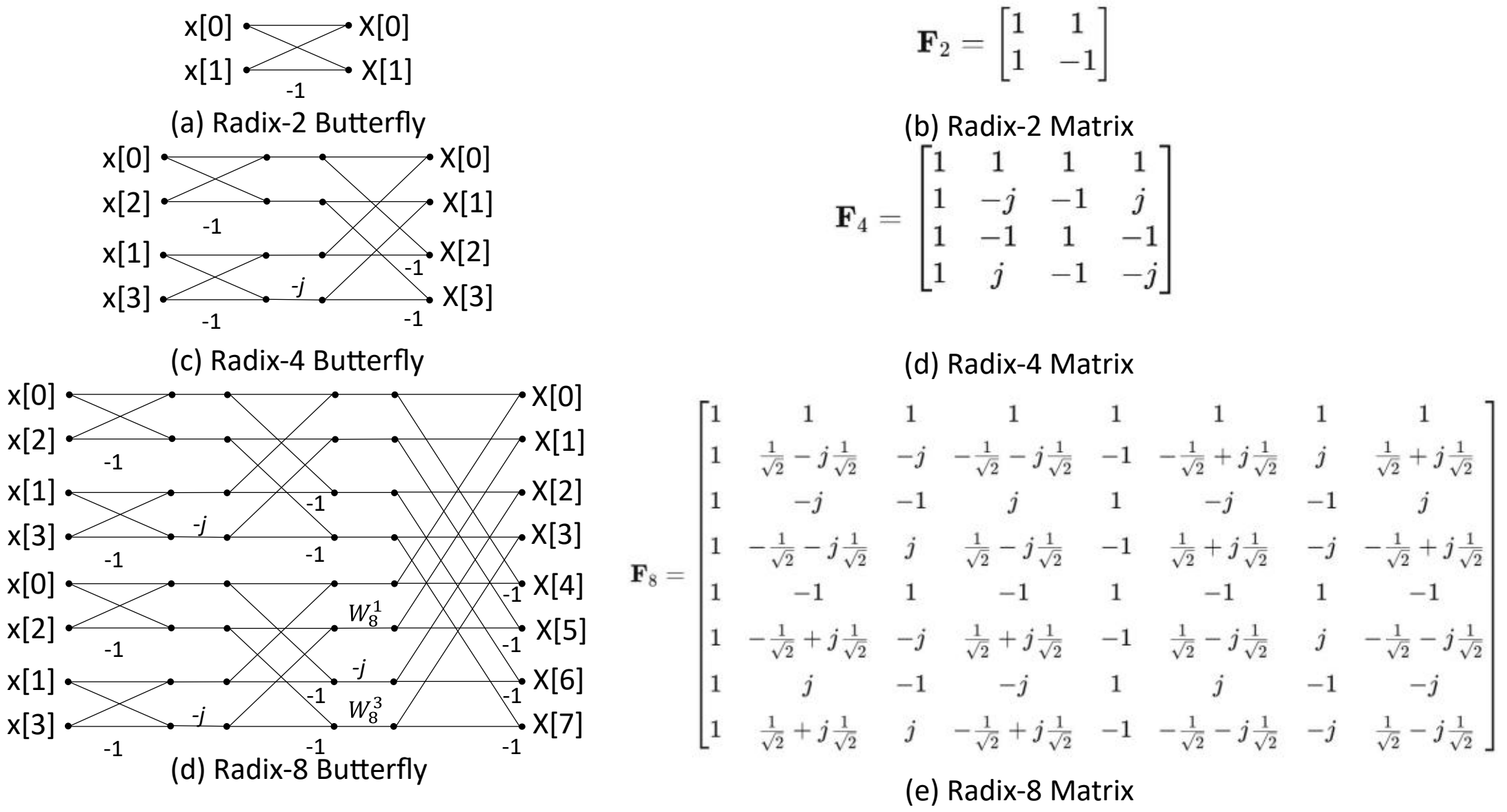

**Fig. 1 Radix-n Butterfly Diagrams and Matrices.**

### 2.2.2 Approximate DFT (ADFT)

Even though radix-*n* FFT reduces the computational complexity to $O(N\log_2 N)$, many applications still demand even faster implementations, such as real-time applications [21, 22]. To further reduce the computational cost and achieve a more efficient FFT process, Approximate DFT (ADFT) is developed [23, 24]. The core idea of ADFT is that the radix-*n* FFT can be further improved as matrices of Radix-*n* are decomposable, and many elements of the DFT matrix are integers $0, \pm 1, \pm j$, which makes computing much faster compared to using complex matrix elements.. Cintra et.al developed 8-, 16-, and 32-point DFT approximations to replace exact DFT building blocks in FFT computations. Madanayake et.al [23] implemented these algorithms in hardware for RF beamforming applications and evaluated their performance in terms of signal-to-noise ratio gains.

### 2.3 Sparse Fourier Transform (SFT)

SFT is another simplified algorithm of DFT, which utilizes the "sparsity" of the signal: only a small part of the points in the frequency domain has large values, while the values of the majority of the other points tend to be close to zero [25, 26]. In communication engineering, the output of the DFT is sparse or approximately sparse [27]. One can take this advantage to devise an "output-sensitive" technology to avoid computing all outputs of the frequency domain. Instead of calculating all the magnitudes of the frequency domain, the SFT focuses on the computed large coefficients. Therefore, SFT is suitable for coding communication signals, such as sounds and text, as they have a detectable range of frequencies. That is to say, SFT does not work for all cases, especially when the frequencies of the signal are complicated [28].

## 3. Sparse Integer Fourier Transform (SIFT)

In this section, we formulate the approximation of the DFT matrix as a design optimization problem, utilizing a matrix decomposition method to reduce the number of design variables. Then, we proposed two methods to approximate twiddle factors using optimization algorithms and reconstructed a diagonal matrix from the matrix decomposition. We compared the two methods and selected the more accurate one. By combining optimization algorithms, the twiddle factors are efficiently expressed as integer complex elements.

### 3.1 Decomposition of the DFT matrix

In order to further reduce the number of design variables in the optimization process, a feasible idea is to decompose the DFT matrix into several integer matrices. The whole approximation can be expressed by the following optimization function:

$$\begin{aligned} &\min : \mathrm{RMSE}(\boldsymbol{\mathcal{A}}_n, \beta_n, K_n) = \sqrt{\| \beta \mathbf{F}_{2^n} - \prod_{i=1}^{K} \mathbf{A}_{2^n}^{i} \|}, \\ &s.t.: \mathbf{A}_{2^n}^{i}[l, m] \in \ x + jy \mid x, y \in Z \end{aligned} \tag{4}$$

The $2^N$-point DFT matrix is decomposed into several matrices. Fig. 2 shows the framework containing two components, DFT matrix decomposition and FT matrix fitting:

**DFT matrix decomposition.** The DFT matrix is decomposed into four sparse matrices at this stage to simplify these multiplications, and two of them, $J_{2^n}$ and $P_{2^n}$, are integer matrices. However, the diagonal matrix contains elements such as $W_{2^N}^1, W_{2^N}^2, \ldots, W_{2^N}^{2^{N-1}-1}$, which are difficult to approximate by integers. The details are shown in Section 3.2.

**DFT matrix fitting.** The matrix that should be approximated is a diagonal matrix $D_{2^n}$ with $2^n$-1 complex numbers. The Differential Evolution algorithm is applied to approximate $D_{2^n}$ through $K$ integer diagonal matrices $a_1, a_2, \ldots, a_K$.

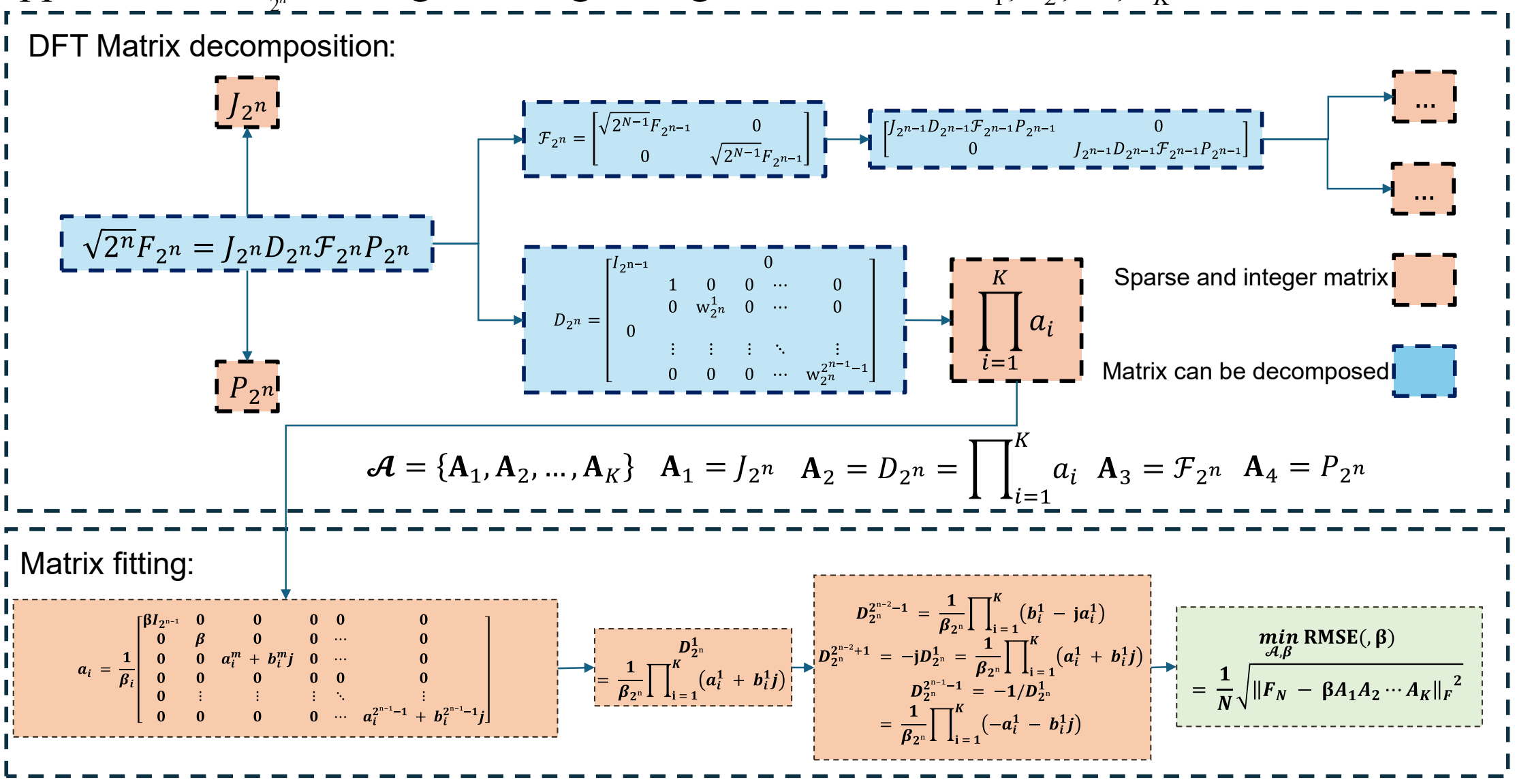

**Fig. 2 Framework of DFT approximation.**

Most elements of the DFT matrix are made of complex and floating-point numbers. Therefore, decomposing the DFT matrix into sparse and integer matrices could reduce the computational complexity and simplify the multiplications.

### 3.1.1 Motivation for matrix decomposition

Although the DFT matrix can be approximated directly without factorization, we adopt matrix decomposition to reduce the number of optimization variables. By retaining the known sparse integer factors and approximating only the remaining diagonal coefficients, the proposed approach reduces the dimension of the optimization problem. The following part provides a plausible method to fit the DFT matrix forcibly: We plan to reconstruct the DFT matrix using several multiplications of sparse and integer matrices. It is plausible to assume that all matrices are symmetric, as the radix-$n$ DFT matrix is symmetric.

Assume that all symmetric, sparse, and integer matrices $\boldsymbol{A}_{2^n}^i$ in Eq. can be represented as $\boldsymbol{A}_{2^n}^i = \boldsymbol{B}_{2^n}^i + \boldsymbol{B}_{2^n}^{i\ T}$. Then, we can assume that

$$\boldsymbol{B}_{2^n}^i = \begin{bmatrix} b_{11}^i & 0 & \cdots & 0 \\ 0 & b_{22}^i & \cdots & 0 \\ \vdots & \vdots & \ddots & \vdots \\ 0 & 0 & \cdots & b_{nn}^i \end{bmatrix} + \begin{bmatrix} 0 & b_{jk}^i & \cdots & 0 \\ 0 & 0 & b_{j+1,k+1}^i & 0 \\ \vdots & \vdots & \ddots & \vdots \\ 0 & 0 & \cdots & 0 \end{bmatrix}, \tag{5}$$

where the elements in matrix $b_{jk}^i = \boldsymbol{B}_{2^n}^i[l,m] \in \{x + jy \mid x, y \in Z\}, j < k, i = 1, 2, \ldots, K$ is expressed in the complex form (They need to be solved by optimization algorithms). We are supposed to find all elements in $\boldsymbol{B}_{2^n}^i$. However, in each matrix, there are $2(2^n + 2^n - 1)$ variables so $K$ matrices need $2K(2^n + 2^n - 1)$ variables.

In optimization problems, the search space is so large with plenty of locally optimal solutions that none of the optimization algorithms can cope with this problem. Therefore, the periodicity and symmetry properties of the DFT matrix and some decomposition methods provide great benefits to minimizing the search space of the optimization process.

### 3.1.2 The decomposition method

According to the property of the DFT matrix, it can be decomposed into the following four sparse matrices [29]. Thus, we decompose it into 4 sparse matrices through the existing recursion:

$$\mathbf{F}_{2^n} = \begin{bmatrix} \mathbf{I}_{2^{n-1}} & \mathbf{I}_{2^{n-1}} \\ \mathbf{I}_{2^{n-1}} & -\mathbf{I}_{2^{n-1}} \end{bmatrix} \begin{bmatrix} \mathbf{I}_{2^{n-1}} & \mathbf{O} \\ \mathbf{O} & \mathbf{d}_{2^{n-1}} \end{bmatrix} \begin{bmatrix} \mathbf{F}_{2^{n-1}} & \mathbf{O} \\ \mathbf{O} & \mathbf{F}_{2^{n-1}} \end{bmatrix} \begin{bmatrix} \{e_{2k}\}_{k=0,\cdots,n/2-1} \\ \{e_{2k+1}\}_{k=0,\cdots,n/2-1} \end{bmatrix} = \prod_{K=1}^{4} \mathbf{A}_{2^n}^K \tag{6}$$

where the diagonal matrix $\mathbf{D}_{2^n-1}$ is

$$\mathbf{D}_{2^n-1} = \mathrm{diag}([1, \cdots, W_{2^n}^{2^{n-1}-2}, W_{2^n}^{2^{n-1}-1}]) = \begin{bmatrix} \mathbf{I}_{2^{n-1}} & \mathbf{O} \\ \mathbf{O} & \mathbf{d}_{2^{n-1}} \end{bmatrix} \tag{7}$$

Let $\mathbf{A}^1_{2^n-1} = \begin{bmatrix} \mathbf{I}_{2^n-1} & \mathbf{I}_{2^n-1} \\ \mathbf{I}_{2^n-1} & \mathbf{-I}_{2^n-1} \end{bmatrix}$, $\mathbf{A}^2_{2^n} = \begin{bmatrix} \mathbf{I}_{2^{n-1}} & \mathbf{O} \\ \mathbf{O} & \mathbf{d}_{2^{n-1}} \end{bmatrix}$, $\mathbf{A}^3_{2^n} = \begin{bmatrix} \mathbf{F}_{2^{n-1}} & \mathbf{O} \\ \mathbf{O} & \mathbf{F}_{2^{n-1}} \end{bmatrix}$, $\mathbf{A}^4_{2^n} = \begin{bmatrix} \{e_{2k}\}_{k=0}^{n/2-1} \\ \{e_{2k+1}\}_{k=0}^{n/2-1} \end{bmatrix}$, and $e_n$ is a unit vector with its $n^{th}$ element being 1.

Therefore, the $2^n$ matrix is decomposed into sparse matrices which are suitable for the approximation through optimization algorithms.

### 3.2 Approximation of the diagonal matrix $A^3_{2^n}$

In this section, we will approximate the complex element $W^k_{2^n}$, as known as the twiddle factor, of the diagonal matrix $\mathbf{d}_{2^n}$ in Eq. (7), using the following expression

$$W^k_{2^n} = \frac{1}{\beta^k_{2^n}} \prod_{i=1}^{m} (a^{ki}_{2^n} + j b^{ki}_{2^n}), \tag{8}$$

where we set $a_i, b_i = [0, \pm 1, \pm 2^1, \pm 2^2, \pm 2^3]$ in small ranges to save computational power. When approximating the diagonal matrix, we proposed two methods. We compared the accuracy of the two methods in Section 3.1.

The first method (Method I) is to use the periodicity and symmetry to approximate the diagonal matrix, then reconstruct the radix-*n* matrices. In the diagonal matrix $\mathbf{d}_2$ of radix-4, there is no need to approximate since $W^1_4 = -j$. For the diagonal matrix $\mathbf{d}_4$ in radix-8, we are only required to approximate the term $W^1_8 = \frac{\sqrt{2}}{2} - \frac{\sqrt{2}}{2} j = \frac{1}{\beta^1_8} \prod_{i=1}^{m} (a^{1i}_8 - j b^{1i}_8)$, as other terms can be expressed using the following formula $W^2_8 = -j, W^3_8 = W^2_8 W^1_8 = -\frac{j}{\beta_8} \prod_{i=1}^{m} (a^{1i}_8 - j b^{1i}_8)$. Therefore, the diagonal matrix $\mathbf{d}_4$ is

$$\mathbf{d}_4 = \mathrm{diag}([1, \frac{1}{\beta^1_8} \prod_{i=1}^{m} (a^{1i}_8 - j b^{1i}_8), -j, -\frac{j}{\beta_8} \prod_{i=1}^{m} (a^{1i}_8 - j b^{1i}_8)]) \tag{9}$$

For the diagonal matrix $\mathbf{d}_8$ in radix-16, we need to approximate the twiddle-factor $W^1_{16}$ using Eq. (8). Therefore, we have

$$\mathbf{d}_8 = \text{diag}([1, W_{16}^1, W_8^1, \frac{1}{\beta_{16}}\prod_{i=1}^{m}(b_{16}^{1i} - ja_{16}^{1i}), -j, \frac{-j}{\beta_{16}}\prod_{i=1}^{m}(a_{16}^{1i} - jb_{16}^{1i}),$$
$$-\frac{j}{\beta_8}\prod_{i=1}^{m}(a_8^{1i} - jb_8^{1i}), \frac{1}{\beta_{16}}\prod_{i=1}^{m}(-a_{16}^{1i} - jb_{16}^{1i})]) \quad (10)$$

In radix-32, we have more terms to approximate. Assuming $W_{32}^1 = \frac{1}{\beta_{32}^1}\prod_{i=1}^{m}(a_{32}^{1i} - jb_{32}^{1i})$, and $W_{32}^3 = \frac{1}{\beta_{32}^3}\prod_{i=1}^{m}(a_{32}^{3i} - jb_{32}^{3i})$, we can represent $\mathbf{d}_{16}$ as

$$\mathbf{d}_{16} = \text{diag}([1, W_{32}^1, W_{16}^1, W_8^3, W_8^1, \frac{1}{\beta_{32}^3}\prod_{i=1}^{m}(b_{32}^{3i} - ja_{32}^{3i}), \frac{1}{\beta_{16}}\prod_{i=1}^{m}(b_{16}^{1i} - ja_{16}^{1i}),$$
$$\frac{1}{\beta_{32}^1}\prod_{i=1}^{m}(b_{32}^{1i} - ja_{32}^{1i}), -j, -\frac{j}{\beta_{32}^1}\prod_{i=1}^{m}(-a_{32}^{1i} - jb_{32}^{1i}), \frac{-j}{\beta_{16}}\prod_{i=1}^{m}(a_{16}^{1i} - jb_{16}^{1i}),$$
$$-\frac{j}{\beta_{32}^3}\prod_{i=1}^{m}(-a_{32}^{3i} - jb_{32}^{3i}), -\frac{j}{\beta_8}\prod_{i=1}^{m}(a_8^{1i} - jb_8^{1i}), \frac{1}{\beta_{32}^3}\prod_{i=1}^{m}(b_{32}^{3i} - ja_{32}^{3i}),$$
$$\frac{1}{\beta_{16}}\prod_{i=1}^{m}(-a_{16}^{1i} - jb_{16}^{1i}), \frac{1}{\beta_{32}^1}\prod_{i=1}^{m}(-a_{32}^{1i} - jb_{32}^{1i})]) \quad (11)$$

There are 4 twiddle factors needs approximation in Method I.

Another method (Method II) uses the power of the first twiddle factor $W_{2^n}^1$ to approximate remaining twiddle factors, i.e.,

$$W_{2^n}^i = (W_{2^n}^1)^i \quad (12)$$

There are 3 twiddle factors that need approximation in this method.

### 3.3 Reconstruction of DFT matrix

In the stage of matrix decomposition, three of the decomposed matrices are available: $\boldsymbol{A}_{2^n}^1$ and $\boldsymbol{A}_{2^n}^2$ can be expressed by known quantities while $\boldsymbol{A}_{2^n}^3$ can be represented through $\boldsymbol{F}_{2^{n-1}}$ figured by the latest step. Hence, only matrix $\boldsymbol{A}_{2^n}^3$ needs approximation. Therefore, we need to solve the twiddle factors $W_{2^n}^k$ of $\boldsymbol{A}_{2^n}^3$ and the optimization problem is defined below:

$$\min_{a_{2^n}^{ki}, b_{2^n}^{ki}, \beta_{2^n}^{k}} :\| W_{2^n}^{k} - \frac{1}{\beta_{2^n}^{k}} \prod_{i=1}^{m} (a_{2^n}^{ki} + j b_{2^n}^{ki}) \|, \tag{13}$$
$$\{a_{2^n}^{ki}, b_{2^n}^{ki}\} \in \{0, \pm 2^0, \pm 2^1, \pm 2^2, \pm 2^3\}, \beta_{2^n}^{k} \in (1, 10^6)$$

where $m$ denotes the number of parameters that participate in this approximation. We should reduce the number of parameters to minimize the multiplication process.

We apply some optimization algorithms to approximate $\boldsymbol{A}_{2^n}^{3}$ through $\prod_{i=1}^{K} a_i$ which can be solved by optimization algorithms such as Differential Evolution (DE), Genetic Algorithm (GA), NSGAII, and NSGAIII. We compare the performance of these algorithms. Genetic Algorithm and NSGAII outperform other algorithms, so we use their results as the approximated twiddle factors.

After approximation, the required twiddle factors are represented as

$$\begin{cases} \hat{W}_8^1 = \frac{1}{181}(-2j)^3(2-2j)^3, \\ \hat{W}_{16}^1 = \frac{1}{325}(-j)(1-8j)^2(1-2j)^2, \\ \hat{W}_{32}^1 = \frac{1}{26107}(-8j)^2(1-8j)(4-8j)(4-4j), \\ \hat{W}_{32}^3 = \frac{1}{289}(-j)(1-4j)^4 \end{cases} \tag{14}$$

Table 1 shows the errors in the approximation of twiddle factors. These errors are at a low level and will be utilized to reconstruct the DFT matrices and process the Fourier Transform.

**Table 1 Twiddle Factors of Analytical DFT matrices and Proposed methods.**

| | True value | Approximation | Error of norm |
|---|---|---|---|
| $W_8^1$ | 0.7071067-0.7071067j | $0.7071823 - 0.7071823j$ | 1.068e-4 |
| $W_{16}^1$ | 0.9238795-0.3826834j | $0.9230769 - 0.3846154j$ | 2.092e-3 |
| $W_{32}^1$ | 0.9807853-0.1905903j | $0.9805799 - 0.1961160j$ | 1.046e-3 |
| $W_{32}^3$ | 0.8314696-0.5555702j | $0.8304498 - 0.5570934j$ | 1.833e-3 |

Finally, we use the solved twiddle factors to reconstruct the DFT matrix. As radix-2 and radix-4 DFT matrices are not required to be approximated, we will rebuild the DFT matrices of radix-8, radix-16, and radix-32 in this section, using Eq. (6).

# 4. Theoretical Results and Analysis

In this section, we present the results of these decomposed matrices using SIFT and evaluate the error with the exact FFT after reconstruction. Then, the computational cost is also evaluated.

### 4.1 Decomposed matrices

The three matrices of radix-8, radix-16, and radix-32 are shown as follows. As radix-2 and radix-4 are exact solutions, there is no need for approximation.

For radix-8, from Eq. (6) in section 3.1.2, the reconstructed $\mathbf{F}_8$

$$\mathbf{F}_8 = \begin{bmatrix} \mathbf{I}_4 & \mathbf{I}_4 \\ \mathbf{I}_4 & \mathbf{-I}_4 \end{bmatrix} \mathbf{A}_8^2 \begin{bmatrix} \mathbf{F}_4 & \mathbf{O} \\ \mathbf{O} & \mathbf{F}_4 \end{bmatrix} \begin{bmatrix} \{e_{2k}\}_{k=0,1,2,3} \\ \{e_{2k+1}\}_{k=0,1,2,3} \end{bmatrix}, \tag{15}$$

where

$$\mathbf{A}_8^2 = \frac{1}{181} \begin{bmatrix} 181\mathbf{I}_4 & \\ & \mathrm{diag}([181,1,181,1]) \end{bmatrix} \begin{bmatrix} \mathbf{I}_4 & \\ & \mathrm{diag}([1,-2j,-j,2]) \end{bmatrix} \begin{bmatrix} \mathbf{I}_4 & \\ & \mathrm{diag}([1,-2j,-j,2]) \end{bmatrix}^2 \begin{bmatrix} \mathbf{I}_4 & \\ & \mathrm{diag}([1,-2j,1,-2j] \end{bmatrix}^3 \tag{16}$$

For radix-16, the diagonal matrix can be approximated by

$$\mathbf{F}_{16} = \begin{bmatrix} \mathbf{I}_8 & \mathbf{I}_8 \\ \mathbf{I}_8 & \mathbf{-I}_8 \end{bmatrix} \mathbf{A}_{16}^2 \begin{bmatrix} \mathbf{F}_8 & \mathbf{O} \\ \mathbf{O} & \mathbf{F}_8 \end{bmatrix} \begin{bmatrix} \{e_{2k}\}_{k=0,\cdots,7} \\ \{e_{2k+1}\}_{k=0,\cdots,7} \end{bmatrix}, \tag{17}$$

where the diagonal matrix is shown as follows

$$\mathbf{A}_{16}^2 = \prod_{i=1}^{7} \begin{bmatrix} \mathbf{I}_8 & \mathbf{O} \\ \mathbf{O} & \mathbf{a}_8^i \end{bmatrix} \tag{18}$$

and

$$\begin{cases} \mathbf{a}_8^1 = \mathrm{diag}([1,\frac{1}{325},\frac{1}{181},\frac{1}{325},1,\frac{1}{325},\frac{1}{181},\frac{1}{325}]) \\ \mathbf{a}_8^2 = \mathrm{diag}([1,-j,-2j,-1,-j,-1,-2,-j]) \\ \mathbf{a}_8^3 = \mathrm{diag}([1,1-8j,-2j,8-j,1,1-8j,-2j,-1]), \mathbf{a}_8^4 = \mathbf{a}_8^3 \\ \mathbf{a}_8^5 = \mathrm{diag}([1,1-2j,2-2j,2-j,1,1-2j,2-2j,-1-2j]), \mathbf{a}_8^6 = \mathbf{a}_8^5 \\ \mathbf{a}_8^7 = \mathrm{diag}([1,1,2-2j,1,1,1,2-2j,1]) \end{cases} \tag{19}$$

For radix-32, the diagonal matrix is represented by the same method.

$$\mathbf{F}_{32} = \begin{bmatrix} \mathbf{I}_{16} & \mathbf{I}_{16} \\ \mathbf{I}_{16} & \mathbf{-I}_{16} \end{bmatrix} \mathbf{A}_{32}^{2} \begin{bmatrix} \mathbf{F}_{16} & \mathbf{O} \\ \mathbf{O} & \mathbf{F}_{16} \end{bmatrix} \begin{bmatrix} \{e_{2k}\}_{k=0,\cdots,15} \\ \{e_{2k+1}\}_{k=0,\cdots,15} \end{bmatrix}, \tag{20}$$

where the diagonal matrix is shown as follows

$$\mathbf{A}_{32}^{2} = \frac{1}{v} \prod_{i=1}^{6} \begin{bmatrix} \mathbf{I}_{16} & \mathbf{O} \\ \mathbf{O} & \mathbf{a}_{16}^{i} \end{bmatrix}, \tag{21}$$

and all matrices $\mathbf{a}_{16}^{i}$ and the amplification factor $v$ are demonstrated below

$$\begin{cases} \mathbf{a}_1 = v\,\mathrm{diag}([1, \frac{-8j}{26107}, \frac{-j}{325}, \frac{1-4j}{289}, \frac{-2j}{181}, \frac{4-j}{289}, \frac{1}{325}, \frac{8}{26107}, \\ \quad -j, \frac{-8}{26107}, \frac{-1}{325}, \frac{-j-4}{289}, \frac{-2}{181}, \frac{-4j-1}{289}, \frac{-j}{325}, \frac{-8j}{26107}]) \\ \mathbf{a}_2 = \mathrm{diag}([1, -8j, 1-8j, -j, -2j, 4-j, 8-j, 8, 1, -8j, -1, -j-4, -2j, -4j-1, -1-8j, 8]) \\ \mathbf{a}_3 = \mathrm{diag}([1, 1-8j, 1-8j, 1-4j, -2j, 4-j, 8-j, 8, 1, 1-8j, 1-8j, 1-4j, -2j, 4-j, -1-8j, 8-j]) \\ \mathbf{a}_4 = \mathrm{diag}([1, 4-8j, 1-2j, 1-4j, 2-2j, 4-j, 2-j, 8-4j, 1, 4-8j, 1-2j, 1-4j, 2-2j, 4-j, -1-2j, 8-4j]) \\ \mathbf{a}_5 = \mathrm{diag}([1, 4-4j, 1-2j, 1-4j, 2-2j, 4-j, 2-j, 4-4j, 1, 4-4j, 1-2j, 1-4j, 2-2j, 4-j, -1-2j, 4-4j]) \\ \mathbf{a}_6 = \mathrm{diag}([1, 1, 1, 1, 2-2j, 1, 1, 1, 1, 1, 1, 2-2j, 1, 1, 1, 1]) \\ v = 26107 \times 289 \times 325 \times 181 \end{cases}$$

(22)

### 4.2 Error Analysis of Approximation

After reconstructing the DFT matrix, we apply two criteria to evaluate the performance of approximation: Mean Absolute Percentage Error (MAPE) and Root Mean Squared Error (RMSE). They can be calculated by the following formula:

$$\begin{aligned} MAPE &= \frac{1}{2^n} \left( \sum_{j=1}^{2^n} \sum_{i=1}^{2^n} \frac{|\mathbf{F}_{2^n}(i,j) - \hat{\mathbf{F}}_{2^n}(i,j)|}{\mathbf{F}_{2^n}(i,j)} \right) \times 100\% \\ RMSE &= \sqrt{\frac{1}{2^n} \sum_{j=1}^{2^n} \sum_{i=1}^{2^n} (\hat{\mathbf{F}}_{2^n}(i,j) - \mathbf{F}_{2^n}(i,j))^2} \end{aligned} \tag{23}$$

The MAPE and RMSE of the two proposed methods for radix-$n$ in DFT matrices approximation are shown in Table 2. We can see that both proposed methods are accurate, as the errors of approximation are marginal. The maximal error occurs in the Radix-32 matrix, with the RMSE and MAPE being 2.43e-5 and 0.341%, respectively. Comparing the two methods, Method I outperforms the other. The reason is that errors accumulate in Method II as the power $i$ grows. This defect is avoided in Method I, which utilizes periodicity and symmetry to approximate all twiddle factors. Therefore, we use Method I in the following matrix reconstruction.

**Table 2 Error of DFT matrices approximation of the proposed method I.**

| | MAPE (Method I) | RMSE (Method I) | MAPE (Method II) | RMSE (Method II) |
|---|---|---|---|---|
| Radix-2 | 0 | 0 | 0 | 0 |
| Radix-4 | 0 | 0 | 0 | 0 |
| Radix-8 | 0.00277% | 2.85e-9 | 0.00277% | 2.85e-9 |
| Radix-16 | 0.0550% | 1.09e-6 | 0.158% | 7.66e-6 |
| Radix-32 | 0.204% | 1.01e-5 | 0.341% | 2.43e-5 |

Cintra [30] described the construction of a 32-point approximate DFT and analyzed its arithmetic complexity. A Fast Radix-32 Approximate DFT uses a multiplier-reduced approximation of the radix-32 DFT building block, where most coefficients are restricted to simple values (e.g., $\pm 1,\ \pm j$) so that the implementation can be realized primarily with adders and sign changes, targeting RF beamforming applications. However, a purely multiplier-reduced design (using only simple values) generally increases the approximation error. Compared to their method, the proposed method reduces the MAPE and RMSE of the radix-32 matrix by 201.62 and 7634.7 folds, respectively (i.e., from 41.13% to 0.204% for MAPE, and from 0.7711 to 1.01e-5 for RMSE).

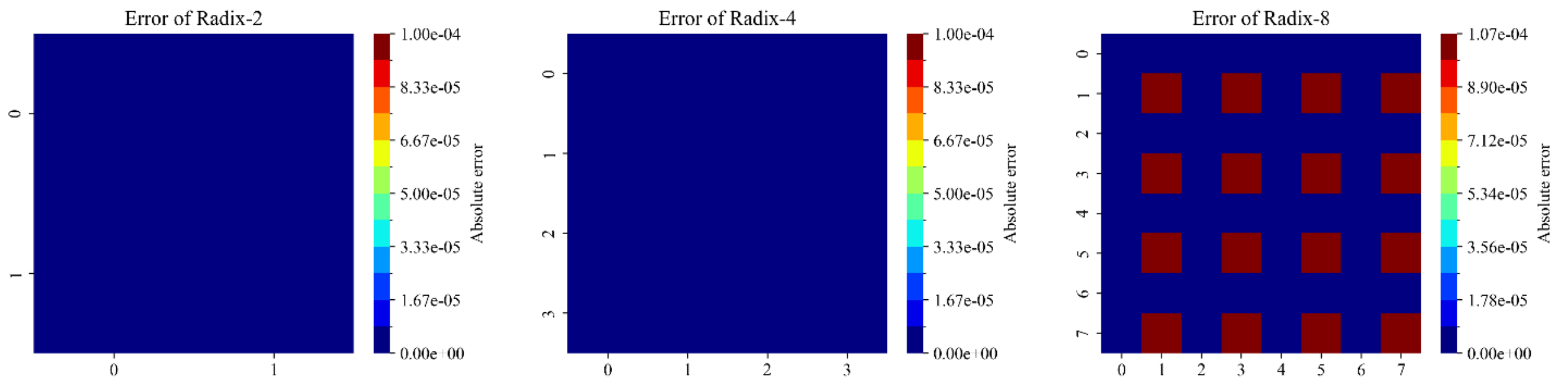

**(a) Heatmap of Radix2 Error** **(b) Heatmap of Radix4 Error** **(c) Heatmap of Radix16 Error**

**Fig. 3 Heatmap of Radix 2,4, and 8 Approximation**

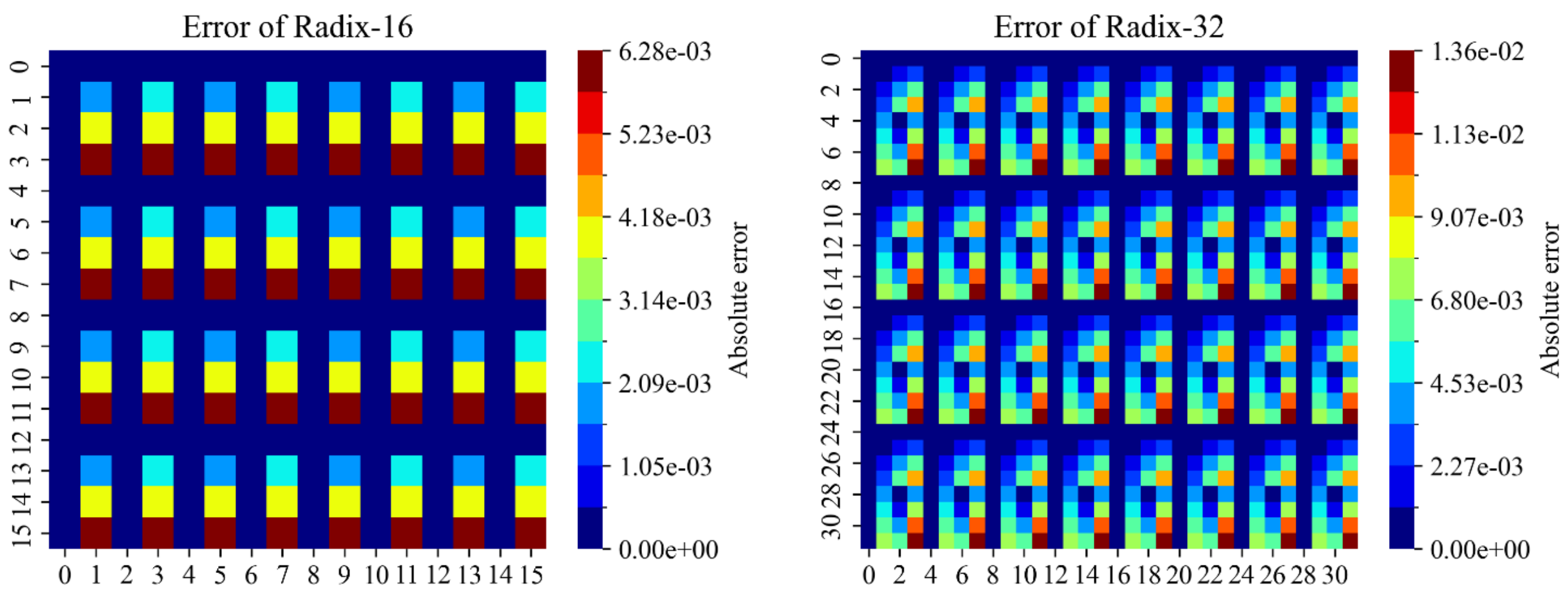

(a) Heatmap of Radix16 Error (b) Heatmap of Radix16 Error

**Fig. 4 Heatmap of Radix 16 and 32 Approximation.**

### 4.3 Computational cost analysis

Similar to implementation-oriented FFT frameworks such as FFTW [31], where algorithm performance is dominated by non-trivial arithmetic operations rather than asymptotic complexity [32], we introduce a simplified hardware cost metric called the computational complexity to facilitate comparison among different decompositions. The hardware complexity is defined as follows

$$C = qL , \tag{24}$$

where $q$ indicates the value range of the elements in the decomposed matrix $\mathbf{A}_k$, and $L$ represents non-trivial complex multiplications. Specifically, the range of the real and imaginary parts of the elements in $\mathbf{A}_k$ is restricted into $\mathcal{P} = \{0, \pm 1, \pm 2, \ldots, \pm 2^{q-1}\}$.

The calculated hardware complexity of the proposed method and the exact FFT are summarized in Table 3. We take $\mathcal{P} = \{0, \pm 1, \pm 2, \pm 4\}$, which corresponds to $q = 3$. Using the above metric, multiplications involving trivial constants such as multiplications with 0, ±1, $\pm j$, and $\pm 1 \pm j$ are excluded, since they can be efficiently implemented using sign changes without dedicated multipliers.

It can be observed from Table 3 that for higher radices, the proposed method consistently requires significantly fewer non-trivial multiplications than the exact FFT. For example, at radix-8, the hardware complexity is reduced from 1024 to 80, corresponding to a reduction of approximately 92%. Similarly, reductions of about 82% and 75% are achieved for radix-16 and radix-32, respectively.

These results indicate that the advantage of the proposed decomposition becomes more pronounced as the radix increases. By constraining the coefficient set and eliminating non-essential multiplications, the proposed method effectively reduces the overall hardware cost while preserving the transformation structure. Therefore, compared with the exact FFT, the hardware complexity of the proposed method is reduced by at least 75% for all considered high-radix cases.

**Table 3 Comparison of the hardware complexity.**

| Radix-n | Proposed Method | FFT |
|---|---|---|
| 2 | 0 | 0 |
| 4 | 0 | 0 |
| 8 | 80 | 1024 |
| 16 | 736 | 4096 |
| 32 | 4032 | 16384 |

# 5. Numerical Experiment

### 5.1 Synthetic function

We use a numerical function combined with several sine and cosine functions to test our proposed method. The test function is shown in Eq. (22).

$$f(t) = 1.2\sin\pi t + 16\cos 5\pi t + 7\cos 13\pi t + 23\sin 18\pi t + 15\sqrt{2}\sin(25\pi t + \frac{\pi}{4}) \quad (25)$$

We use Eq. (3) to analyze the frequencies of the test function. We sampled $N=2^{12}$ points for radix-2, radix-4, radix-8, and radix-16 matrices, and $2^{15}$ for radix-32, since the radix-$n$ method requires $N=n^m$ data points. More data is sampled for radix-32 due to the radix-32 block ($(2^5)^3=2^{15}$).

In our sampling process, the frequency of sampling is 120 Hz, and the time is one period. After the DFT process, we plot the analytical results and the approximated results in the left column of Fig. 5. The errors are shown in the right column plots of Fig. 5. In Table 4, we present the error of the DFT process using the metric of MAPE and RMSE. From the results in Fig. 5 and Table 4, maximal error occurs in the radix-32 method, but still less than 1%. The results show that SIFT can accurately extract frequencies of the test function with marginal errors.

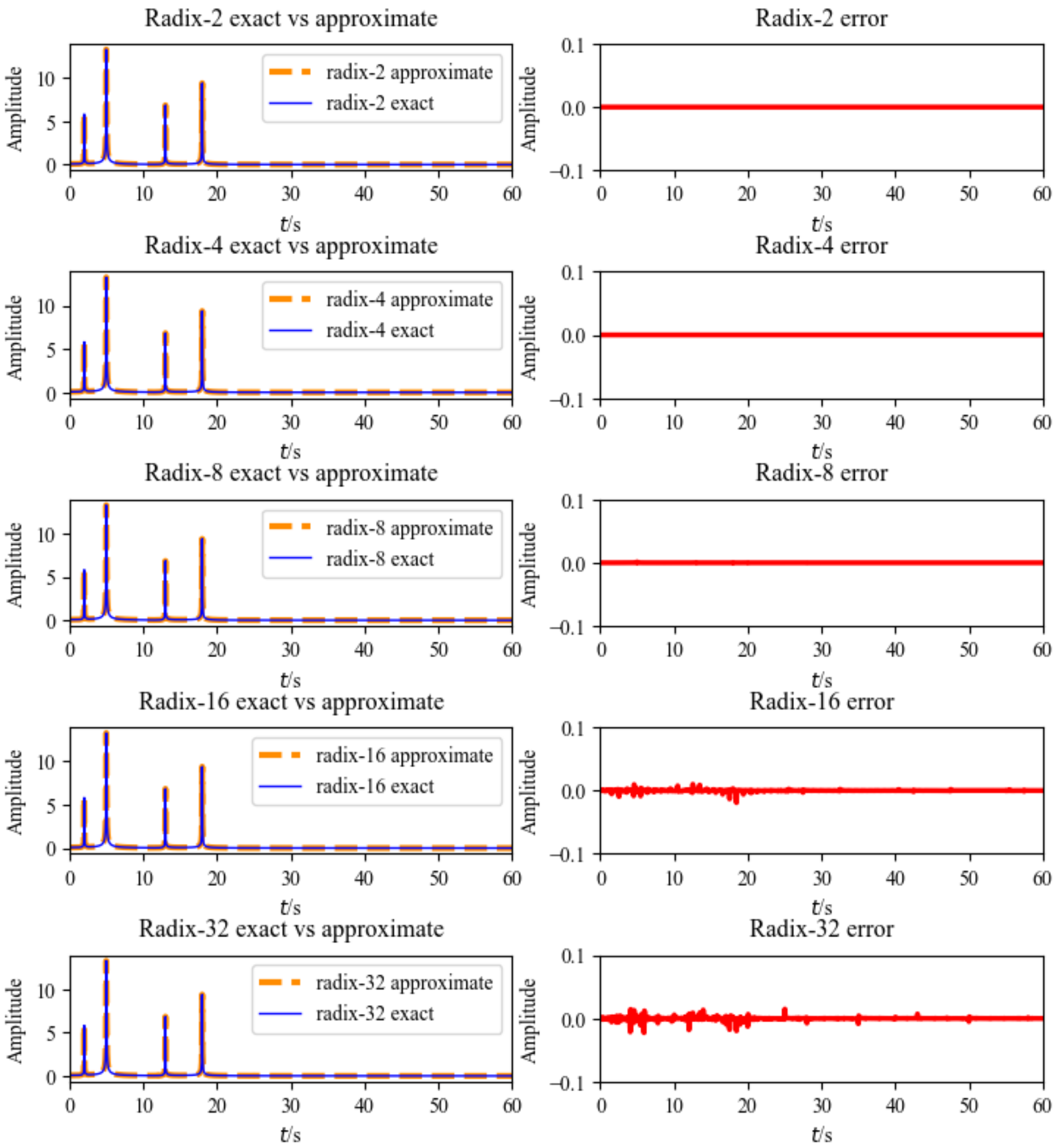

**Fig. 5 Results and errors of Analytical DFT and the proposed method.**

**Table 4 MAPE and RMSE errors of the proposed method.**

| | MAPE | RMSE |
|---|---|---|
| Radix-2 | 0 | 0 |
| Radix-4 | 0 | 0 |
| Radix-8 | 0.04963% | 4.845e-9 |
| Radix-16 | 0.9192% | 8.589e-7 |
| Radix-32 | 2.042% | 6.147e-6 |

## 5.2 Fundamental frequency under the dynamic loads

Structural optimization under time-series loading is critical in many engineering fields, such as civil, mechanical, and aerospace engineering. One of the key components of structural optimization is extracting the fundamental frequency under the dynamic loads, e.g., wind and water. In the design optimization process, we need to calculate the frequencies of the cantilever beam to avoid resonance.

In this example, we consider the design of a cantilever beam (Fig. 7) with a length of 10 m subjected to a dynamic load at the free end. The width and height of the hollowed cantilever beam are 0.1 m and 0.8 m, respectively, made of rubber with a density of 790 kg/$m^3$ and Young's modulus of 2e6 Pa. The thickness of the beam is 0.02 m. The beam needs to bear a dynamic load with a magnitude of 1200 N and a frequency of 5 Hz, $F = 1200\sin(5\pi t)$.

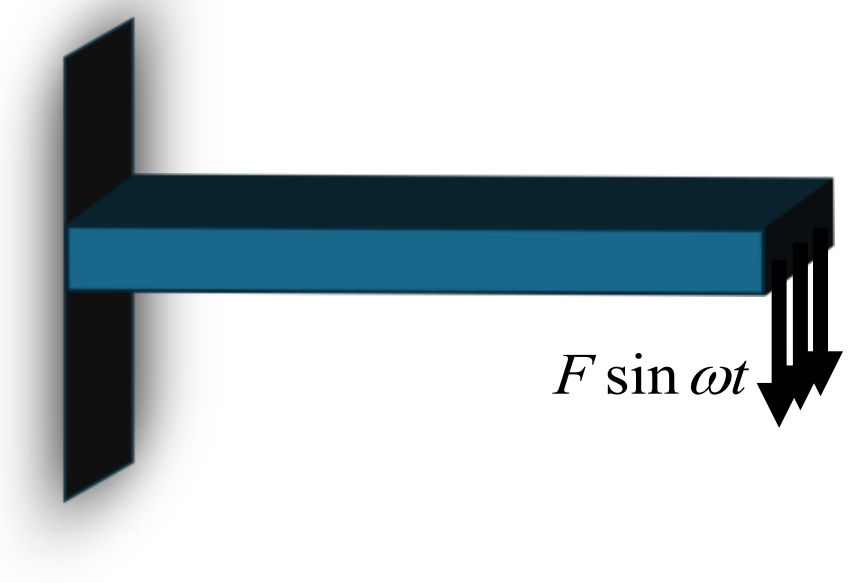

**Fig. 6 Design of a cantilever beam.**

To validate our DFT method, we utilize the Finite Element Method (FEM) for analysis and apply the proposed method to extract the frequency of the load and the fundamental frequency, as well as the other harmonic frequencies. The governing equation is as follows:

$$\frac{\partial^2 y}{\partial x^2}(EI(x)\frac{\partial^2 y}{\partial x^2}) + \rho A(x)\frac{\partial^2 y}{\partial t^2} = \frac{F}{\rho A(x)}\delta(x-L)\sin(\omega t) \tag{26}$$

where the response $y(x,t)$ of the beam is a function of time and deflection, and the boundary conditions are

$$\begin{cases} y(0,t) = 0, \\ y'(0,t) = 0, \\ EI(L)y''(L,t) = 0, \\ EI(L)y^{(3)}(L,t) = \dfrac{F\sin(\omega t)}{\rho A(L)}. \end{cases} \tag{27}$$

as there are no deflection and slope at the fixed end, no shear force at the free end, and the bending moment of the free end is $F\sin(\omega t)/\rho A(L)$.

To solve the governing equation, we decouple the time and displacement first. We can assume the function of deflection to decouple the time and displacement (assuming there is no damping): $y(x,t)=w(x)\sin(\omega t)$, where $w(x)$ is the deflection of the cantilever beam. Then, we substitute it into the governing function, and the item of time can be canceled.

$$\frac{\mathrm{d}^2 w}{\mathrm{d}x^2}(EI(x)\frac{\mathrm{d}^2 w}{\mathrm{d}x^2})-\omega^2\rho A(x)w(x)=\frac{F}{\rho A(x)}\delta(x-L) \tag{28}$$

Subsequently, we generate a mesh with 11 elements using a quartic shape function $\varphi_j^0(x)$ and $\varphi_j^1(x)$ to approximate the deflection $w(x)$, applying FEM. The deflection is represented by

$$w(x)=d_j^0\varphi_j^0(x)+d_j^1\varphi_j^1(x) \tag{29}$$

where $d_j^0$ and $d_j^1$ are the coefficients of the shape functions $\varphi_j^0(x)$ and $\varphi_j^1(x)$, respectively. In this finite element method, we partition the beam into 200 parts, so the mesh includes 200 elements and 201 nodes. Therefore, the response of the beam to the dynamic load is as $y(x,t)=(d_j^0\varphi_j^0(x)+d_j^1\varphi_j^1(x))\sin(\omega t)$.

In addition, the analytical solution for the response can be described by:

$$w(x)=\sum_{i=1}^{\infty}\frac{\phi_i(L)\phi_i(x)}{\omega_i^2(1-(\omega_i/\omega)^2)} \tag{30}$$

where the $\omega_i$ is the natural frequency of the beam and the mode function $\phi_i(x)$ is

$$\phi_i(x)=\cos\beta_i x-\cosh\beta_i x+\frac{\cos\beta_i l+\cosh\beta_i l}{\sin\beta_i l+\sinh\beta_i l}(\sin\beta_i x-\sinh\beta_i x) \tag{31}$$

In the following section, we will use the proposed method to extract the frequency of the dynamic load, and the obtainable frequency of vibration mode.

### 5.2.1 Frequency of the dynamic load

Considering the response of the free end in the time domain, we extract the value

$\omega$ in the following equation,

$$y(L,t) = w(L)\sin(\omega t), \tag{32}$$

where $w(L)$ is approximated by the Finite Element Model.

Using the SIFT, we plot the analytical results, the approximated results (left column in Fig. 9), and the errors (right column in Fig. 9). We can see that the extracted peak in the frequency domain is at 5 Hz, which corresponds to the 5 Hz frequency of the load in this beam problem. Also, the approximation errors are extremely low (maximum error in radix-32 with the RMSE to be $4.21 \times 10^{-9}$), which shows that the proposed method maintains high accuracy in extracting the frequency of the load.

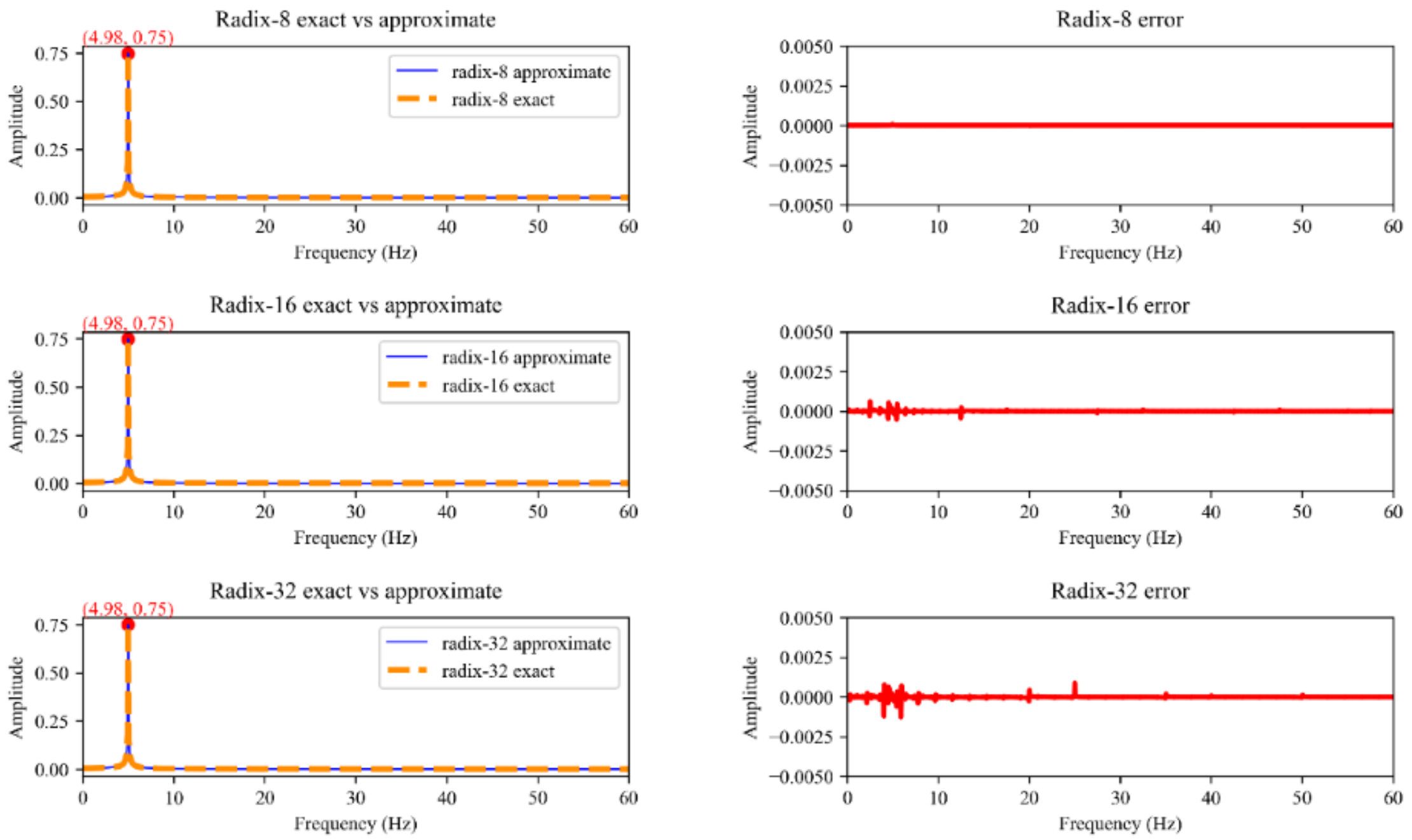

**Fig. 7 Results and errors of the Analytical DFT and the proposed method to extract the frequency of the dynamic load.**

### 5.2.2 Frequency of the modal frequency

In Section 5.2.1, we used the proposed SIFT method to extract the dominant frequency of the external dynamic load from the time-domain response at the free end. Since resonance is governed not only by the load frequency but also by structural modal frequencies, in this section, we establish the beam's modal frequencies using an analytical solution and then compare them with the frequency components extracted from FEM responses.

The solution of the natural angular frequencies for the beam solved by the analytical method is shown below

$$\omega_i = (\frac{\beta_i^2}{L^2})\sqrt{\frac{EI}{\rho A}}, \omega_1 = 0.4341, \omega_2 = 2.7207, \omega_3 = 7.6188, \omega_4 = 14.9302 \qquad (33)$$

where $\beta_n$ is the eigenvalues of the beam and can be obtained by solving the following equation:

$$1+\cos(\beta_n)\cosh(\beta_n) = 0, \beta_1 \approx 1.875, \beta_2 \approx 4.694, \beta_3 \approx 7.855, ..., \beta_i \approx \frac{2i-1}{2}\pi .$$

Using Eq. (29) and (30), we obtain the response of the deflection of the beam using FEM and analytical method, respectively, and plot them in Fig. 8.

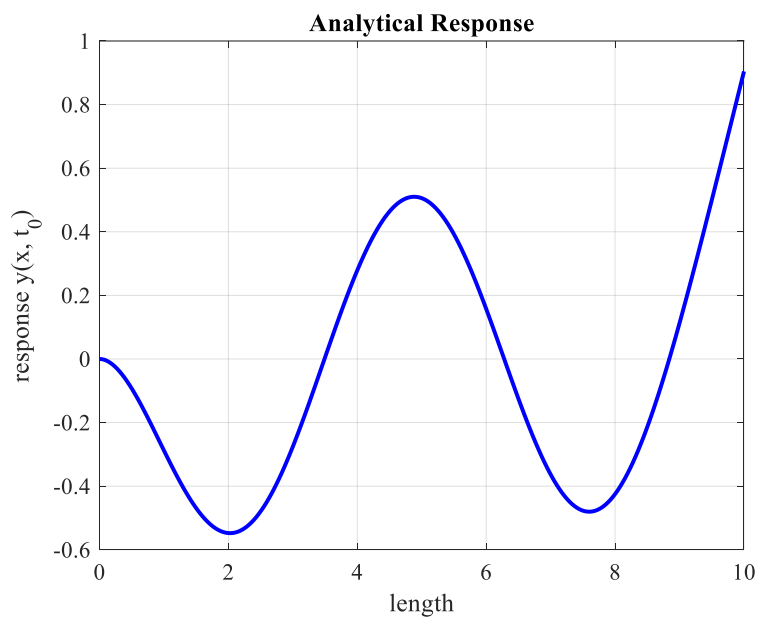

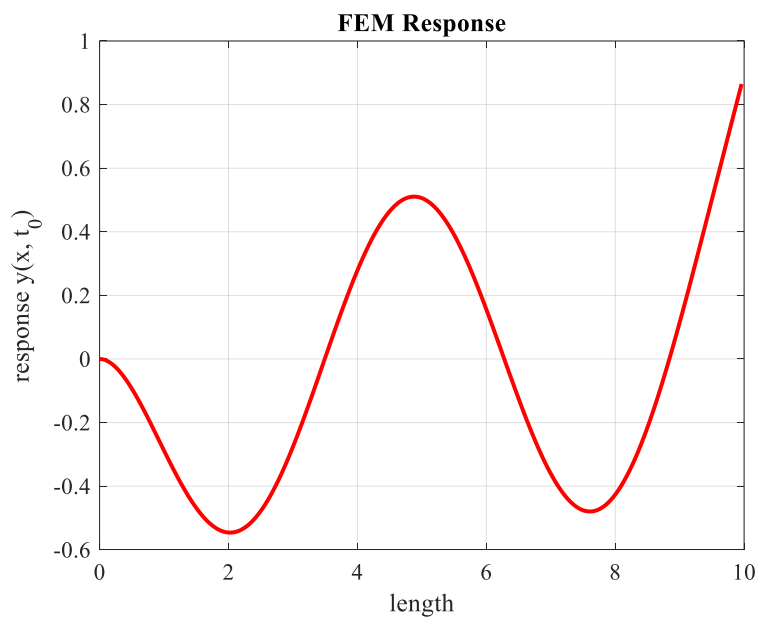

**(a) Analytical Response of the beam** **(b) Response of FEM**

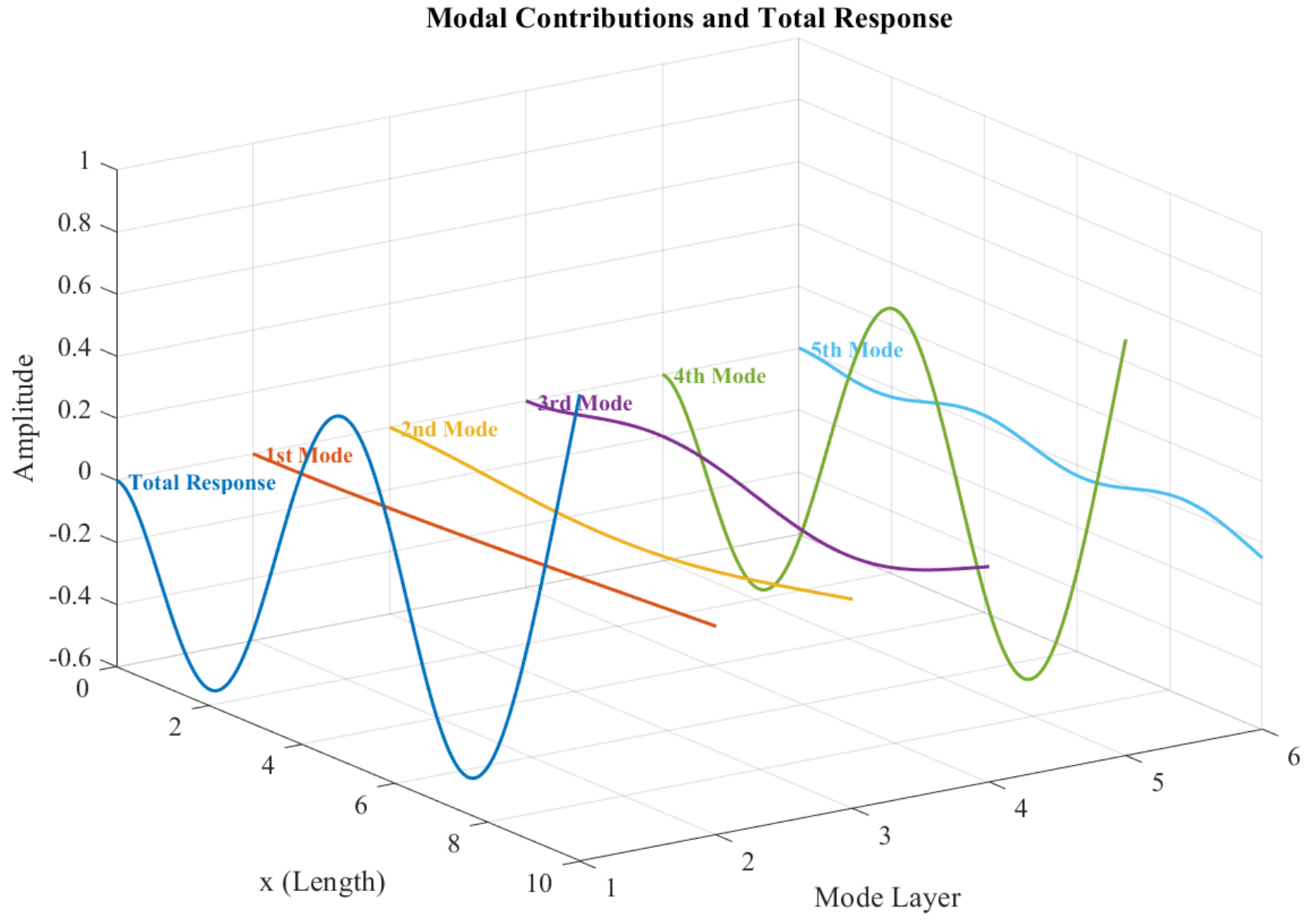

**(c) First 5 vibration modes of the beam**

**Fig. 8 Response of the beam.**

The beam bears the load with an angular frequency of 15.7080 rad/s, which is close to the fourth natural frequency. Thus, the fourth-order mode of vibration is the main vibration mode of this cantilever beam, and the amplitude of the fourth-order vibration mode is significantly increased. The spatial characteristics of the fourth-order mode are 2 complete peak and valley pairs, which means there are two periods in spatial vibration.

As a result, we utilize our approximate radix-$n$ matrices to analyze the beam's vibration and get the main frequencies. We sampled 256 points in responses from analytical and finite element methods. When utilizing the radix-$n$ method, the sampled points must be equal to $2^{(n)m}$, where $m$ is an integer. Since $256=2^8=2^{(1)8}=2^{(2)4}=2^{(4)2}$, 256 points can be applied to radix-2, radix-4, and radix-16 directly. As for radix-8 and radix-32, we use the zero-padding method to enlarge the length of the signal to $2^{(3)3}=512$ and $2^{(5)2}=1024$ points, respectively.

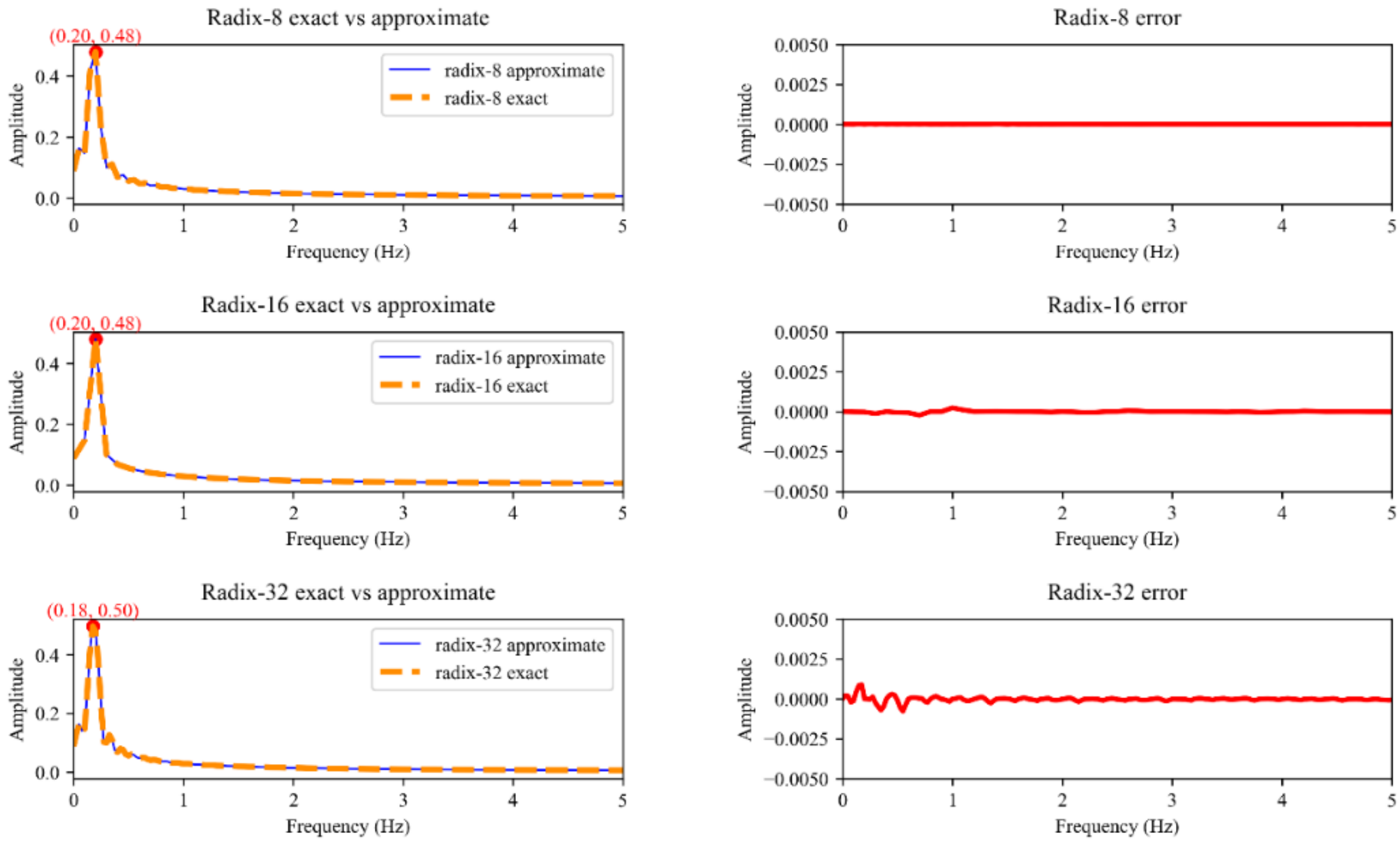

**Fig. 9 Proposed DFT method VS Analytical DFT in Analytical Response of the beam**

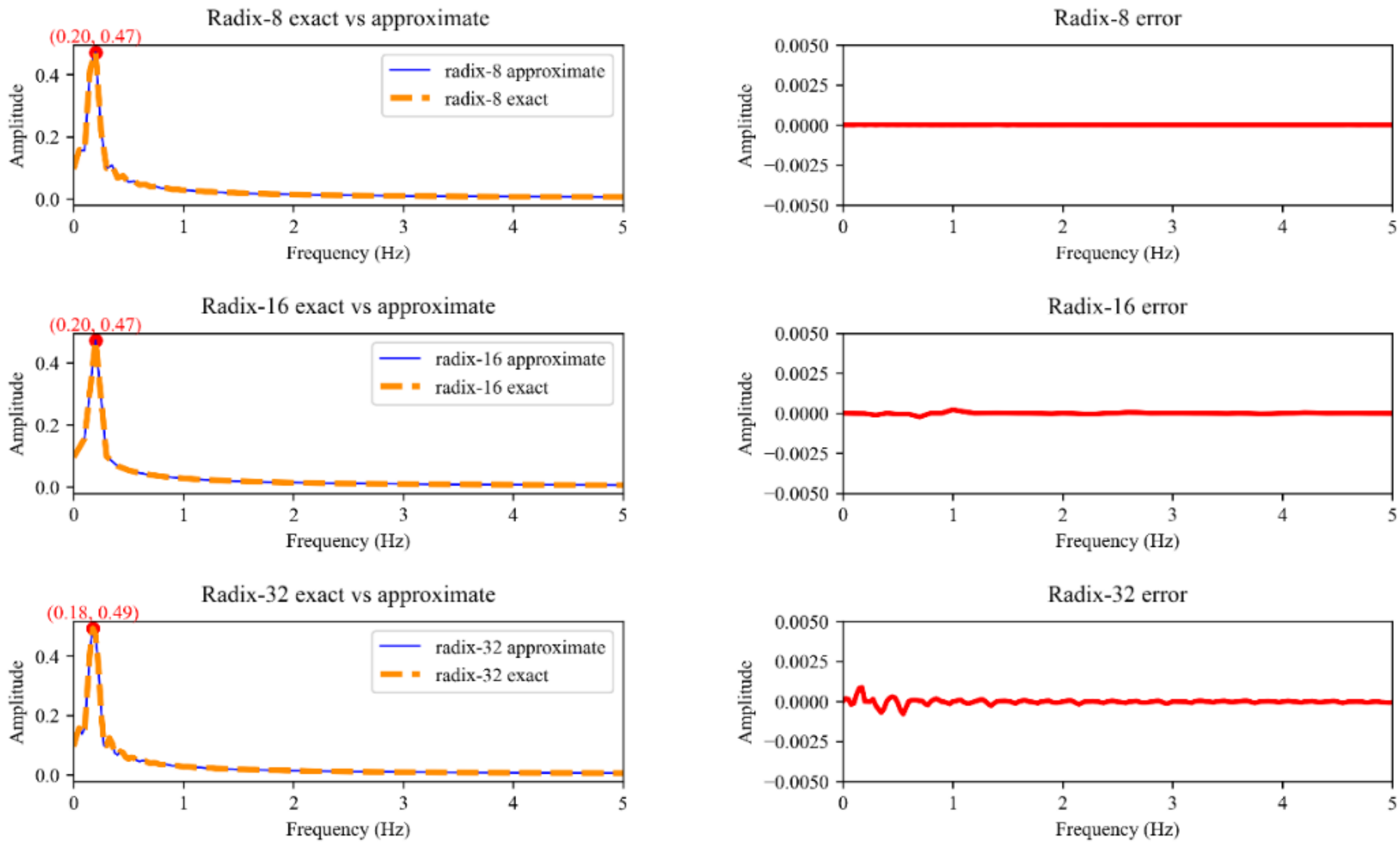

**Fig. 10 Proposed SIFT method VS Analytical DFT in Numerical Response of FEM of the beam.**

The FFT results generated by the analytical deflection and finite element method are shown in Fig. 9 and Fig. 10. When the fourth mode takes charge of the main vibration, there are two valleys and two peaks in the cantilever beam. It indicates two periods in the spatial direction of the beam. The period can be expressed using the fourth eigenvalue of the beam $T_4 = 2\pi/(\beta_4/L) = 5.71046\,\text{m}$. Therefore, the frequency in the spatial direction is $f_4 = 1/T_4 = 0.17501\,\text{m}^{-1}$, which corresponds with the extracted frequency in Fig. 9 and Fig. 10. Some very small errors exist, as there are only about 2 periods in the vibration of the spatial direction.

The SIFT result is more "refined" after zero-padding; the main peak becomes sharper, the side lobes increase, and the resolution becomes higher. The two pictures look identical, as the finite element method approximates the vibration very well. In addition, the maximal error occurs at the radix-32 method, but it stays at a low level (the RMSE of radix-32 result is less than $3.01\times10^{-8}$). The proposed method is more accurate than the test function case, which implies its usefulness in engineering problems. Therefore, the proposed method improves computing efficiency due to the sparse integer operations and preserves the high approximation accuracy.

# 6. Application in Machine Learning

The proposed method SIFT presents high-accuracy FFT results for numerical problems, which makes us consider more interesting applications. In the following case, the approximated FFT output is utilized as the input data for a fault diagnosis problem.

After being validated in Section 5, we present an engineering application of SIFT in time series modeling with machine learning, which is a powerful tool for modeling [33]. Time series data is a common data type in many engineering scenarios, such as signal processing, defect and fault diagnosis, and digital twin technology [34-36]. A key factor for successful time series modeling is how effectively useful patterns can be extracted from noisy data points at a large number of time steps. We tackle this problem using SIFT combined with deep learning.

We use published time series data in the field of fault diagnosis (Condition Monitoring and Fault Diagnosis (CMFD) Dataset) [37] to conduct modeling. This dataset includes 76 groups of time series, with 75 inter-turn short circuit faults (ITSCFs) simulated in the wind turbine generator stator using MATLAB/Simulink, and the remaining one is the data under normal working conditions. Each time series has 319,941 data points, with a total size of 2.8 gigabytes. The above-mentioned data is based on the example: Wind Turbine - MATLAB & Simulink (mathworks.com). In the Simulink WT model, the generator stator is characterized by an equivalent impedance of R=0.0027 Ohm. The 75 (5*5*3) groups of fault segments comprise an increasing series of faulty winding ratios $p_S$:=0.05, 0.1, 0.2, 0.3, 0.5 representing the progression of degradation, a decreasing series of short circuit resistances $R_S$:=1, 0.1, 0.01, 0.001, 0.0001 Ohm models the degradation of the insulation layer, and three phases of ITSCFs.

Each time series is simulated for 80 seconds, using the default wind speed profile from the WT example. The phase-to-phase current (in Amperes) and voltage (in Volts) signals for all three phases are recorded at a sampling frequency of 4 kHz. After plotting the normal data and a group of faulty data in Fig. 11, several differences are observed between the healthy and faulty segments. Since there are about 0.3 million data points per time series, we plot the first 2000 samples, the whole data, and the last 2000 samples.

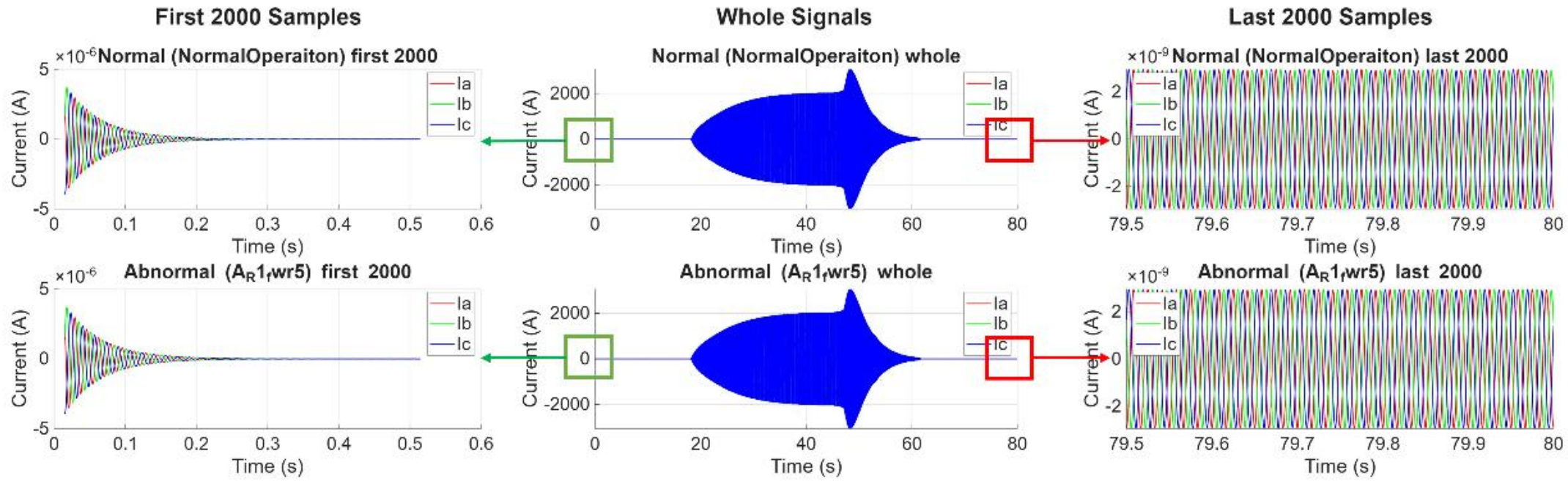

**Fig. 11 Example of the Fault Diagnosis Dataset.**

In Fig. 11, it compares the phase-A current under normal conditions and abnormal conditions. To highlight both the transient and steady-state behaviors, the middle figure shows the full 80 s signal, while the left and right figures zoom into the first 0.5 s and the last 0.5 s segments, respectively. As it shows, the two cases exhibit highly similar waveforms in both the global view and the zoomed-in segments, and their differences are subtle. This indicates that the fault-related signatures are difficult to distinguish by visual inspection in the time domain, motivating the use of spectral feature extraction for reliable fault diagnosis. At first, the whole signal is processed by SIFT, and the frequency of the signal is 60Hz. SIFT also achieves accurate responses compared to the exact FFT. To diagnose the fault categories, the extracted phase and amplitude of signals, including the current and voltage of all three phases, are combined as the input variables. We used an LSTM [38] and a Transformer [39] neural network to classify the signal.

To further demonstrate the practical utility of the proposed FFT/DFT approximation in downstream tasks, we consider a three-class fault-phase classification problem (A/B/C) based on the wind-turbine generator ITSCF dataset. The dataset contains 76 time-series groups (75 faults and one normal case), where each group is generated from a combination of fault severity and phase location, and each time series is simulated for 80 s with a sampling frequency of 4 kHz, resulting in approximately $3.2\times10^5$ samples per signal.

Given the large number of samples per signal, we construct compact feature vectors by uniformly sampling integer time instants from the 80 s record. Specifically, we select 8 representative points at $t$ = 10, 20, …, 80 s for each signal segment.

To diagnose the fault category of the phase, a curve-fitting model is introduced:

$$x(t) = T(t) \times S(t), \tag{34}$$

where we treat the signal as a time series $x(t)$. The time series can be fitted by the extracted trend $T(t)$ and seasonality $S(t)$.

At first, the seasonality of the signal is extracted using SIFT method, and its frequency is $f_0$=60 Hz. Typically, phase drifts exist in many signals. Thus, the seasonality will be fitted as follows:

$$S(t) = \cos(\varphi(t)) = \cos(2\pi f_0 t + \phi(t)), \tag{35}$$

where $\phi(t)$ and $\varphi(t)$ represents the phase drift and current phase, respectively. In particular, for each phase signal, an analytic signal is formed via the Hilbert transform, and the trend of the signal can also be evaluated using the Hilbert transformation.

Firstly, the original signal is shifted by $\pi/2$ degrees as follows

$$\hat{x}^*(t) = T(t)S(t - \frac{\pi}{2}) = T(t)\sin(2\pi f_0 t + \phi(t)) \tag{36}$$

Using the Hilbert Transformation, we get

$$z(t) = \hat{x}(t) + j\hat{x}^*(t) = T(t)(\cos(\varphi(t)) + j\sin(\varphi(t))) = T(t)\exp(j\varphi(t)) \tag{37}$$

where the magnitude of $z(t)$ is the trend

$$T(t) = |z(t)| \tag{38}$$

and the phase also equals the phase of $z(t)$

$$\varphi(t) = \theta(z(t)) \tag{39}$$

As a result, the signal is fitted the Hilbert transformation via the frequency extracted by FFT. Then, two time series classifiers (LSTM and Transformer classifier) are applied to identify the failure.

Two input settings are evaluated. (i) Baseline (raw-signal input): the three-phase current and voltage sequences are directly used as inputs. (ii) Proposed feature input (FFT/Hilbert-based): we first apply the proposed FFT/DFT method to obtain frequency-domain information and then extract physically meaningful amplitude–phase descriptors.

We use the sampled amplitude and phase values at the selected integer time data points ($t$=1, 2, 3, …,80) as the model input features. For the raw signal input with three phases and two measurements (current and voltage), the resulting feature dimension is $80 \times 6 = 480$. For the modeled signal, we input its sampled phase and amplitude in three phases in both current and voltage; the resulting feature dimension is $80 \times 6 \times 2 = 960$. The flowchart for fault diagnosis is shown in Fig. 12.

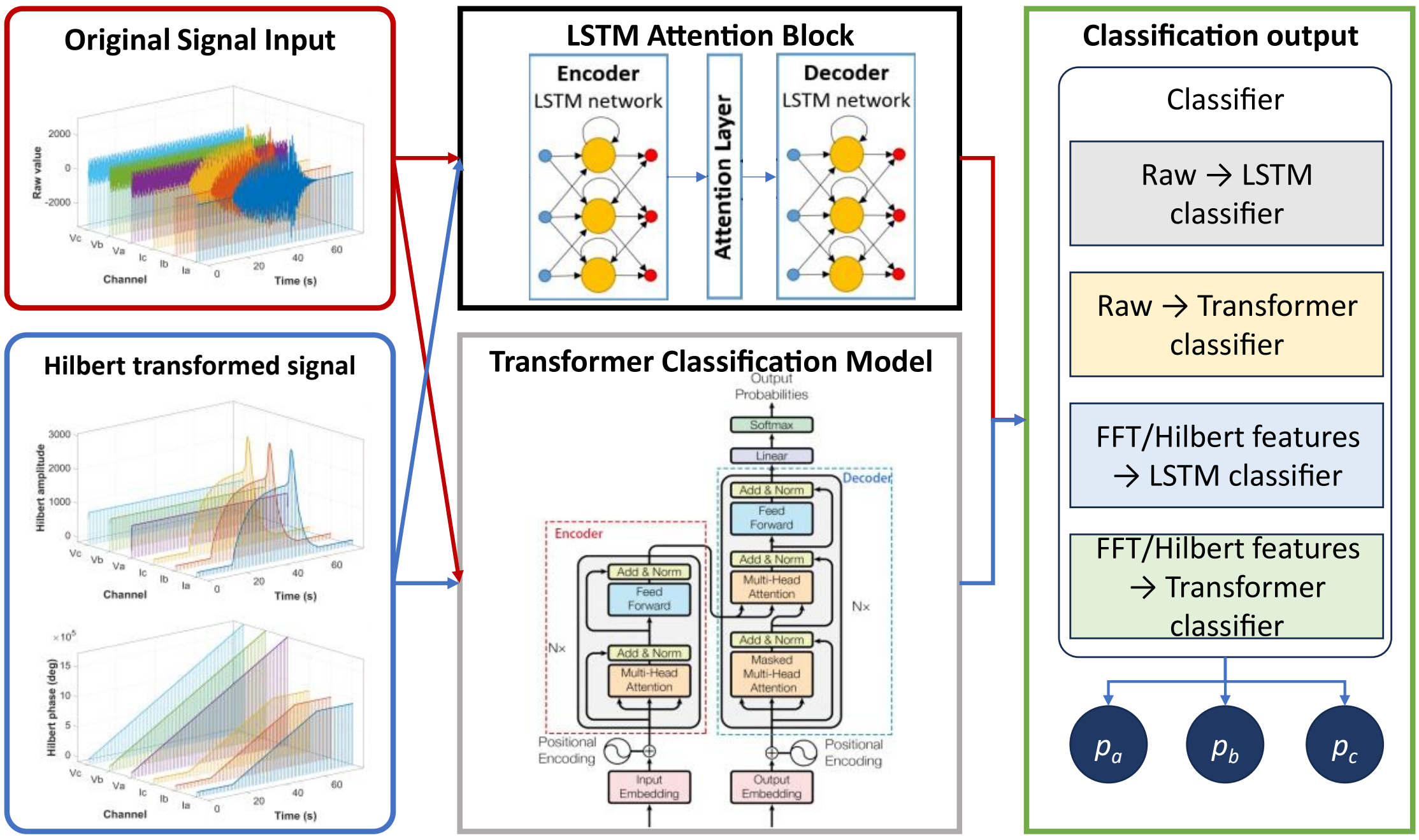

**Fig. 12 Transformer-Based Fault Phase Classification.**

Two sequence models are tested for classification: an LSTM network and a Transformer network. The classification performance is evaluated using 5-fold cross-validation in Table 5.

**Table 5 Accuracy for all models.**

| Model / Fold | 1 | 2 | 3 | 4 | 5 | Mean |
|---|---|---|---|---|---|---|
| LSTM classifier of the raw signal | 0.4333 | 0.5000 | 0.5000 | 0.5333 | 0.6000 | 0.5133 |
| LSTM classifier of the FFT/Hibert Features | 0.7333 | 0.6000 | 0.6667 | 0.7333 | 0.5333 | 0.6533 |
| Transformer classifier of the raw signal | 0.8667 | 0.8000 | 0.7333 | 0.9333 | 0.8000 | 0.8267 |
| Transformer classifier of FFT/Hilbert features | 1.0000 | 1.0000 | 1.0000 | 0.9333 | 0.8000 | 0.9467 |

According to the accuracy evaluated by 4 models in the test dataset, Transformer outperforms LSTM, and classifiers yield higher accuracy with the FFT/Hilbert features. Results show that the Transformer achieves consistently higher accuracy than the LSTM when using the FFT/Hilbert-derived amplitude–phase features, indicating that the proposed frequency-feature extraction preserves the fault information while significantly reducing the input dimensionality compared with directly feeding the full-length time-series signals.

## 7. Conclusion

In this paper, we present a Sparse Integer Fourier Transform (SIFT). Conventional FFTs reduce the computational costs of DFT from O(*N*2) down to O ($N \log_2 N$). SIFT accomplishes further computing savings using sparse and integer matrices as an approximate FFT method. We introduce the optimization method combined with matrix decomposition to reduce the number of design variables to complete our approximations. We propose approximation methods for calculating integer and sparse matrices of radix-2, radix-4, radix-8, radix-16, and radix-32 in the DFT process. Firstly, we decompose these matrices and extract the elements that require approximation. Then, we utilize $2^n$ to approximate a small portion of twiddle factors. The rest of the twiddle factors are approximated using periodicity. This method applies optimization algorithms to solve the approximation factors for twiddle factors in the DFT matrix and maintains higher accuracy. The accuracy of the approximate matrices is validated with a small MAPE and RMSE. The proposed SIFT method is tested in two numerical problems for extracting frequencies. The frequency of the signals is extracted accurately, and the errors are at a low level. Finally, SIFT is applied to a basic fault diagnosis problem. Results also show that SIFT improves classification accuracy for data preprocessing in machine learning.

## Acknowledgments

The authors acknowledge support from the startup fund, the 4-VA grant, and the graduate assistantship provided by Virginia Tech, and the Curriculum Development

Support provided by MathWorks. We would like to express our gratitude to Dr. Robert Canfield for the insightful discussions.

## Conflict of Interest

The authors declared that they have no conflicts of interest with this work.

## Data Availability

The author confirms that all data generated or analyzed during this study are included in this published article.